\documentclass[prd,preprint,superscriptaddress,showpacs,byrevtex]{revtex4}
\usepackage{bm}
\def\fsl#1{\setbox0=\hbox{$#1$}           
   \dimen0=\wd0                                 
   \setbox1=\hbox{/} \dimen1=\wd1               
   \ifdim\dimen0>\dimen1                        
      \rlap{\hbox to \dimen0{\hfil/\hfil}}      
      #1                                        
   \else                                        
      \rlap{\hbox to \dimen1{\hfil$#1$\hfil}}   
      /                                         
   \fi}                                         %
\newcommand{\be}{\begin{equation}}
\newcommand{\ee}{\end{equation}}
\newcommand{\bea}{\begin{eqnarray}}
\newcommand{\eea}{\end{eqnarray}}
\newcommand{\beq}{\begin{equation}}
\newcommand{\eeq}{\end{equation}}
\newcommand{\beqs}{\begin{eqnarray}}
\newcommand{\eeqs}{\end{eqnarray}}

\begin{document}
\title{ 5-Point 1PI Proper Vertex Function of Gluon and the S-Matrix Element }
\author{Gouranga C Nayak }\thanks{G. C. Nayak was affiliated with C. N. Yang Institute for Theoretical Physics in 2004-2007.}
\affiliation{ C. N. Yang Institute for Theoretical Physics, Stony Brook University, Stony Brook NY, 11794-3840 USA}
\date{\today}
\begin{abstract}
As far as the renormalization in perturbative QCD is concerned the n-point one particle irreducible (1PI) proper vertex function is the basic building block where the ultra-violet (UV) divergence occurs when the loop momentum integration limit goes to infinity. In this paper we express the S-matrix element for the $gg \rightarrow ggg$ scattering process at all orders in coupling constant in terms of 5-point, 4-point, 3-point 1PI proper vertex functions and the (full) propagator by using the path integral formulation of QCD.
\end{abstract}
\maketitle
\pagestyle{plain}
\pagenumbering{arabic}
\section{ Introduction }

In quantum field theory one of the main aim is to calculate the scattering cross section which is experimentally measured. The scattering cross section can be calculated from the S-matrix element. In QCD the generating functional in the path integral formulation yields the connected green's function of the parton at all orders in coupling constant. This connected green's function can be used in the LSZ reduction formula to predict the S-matrix element for the partonic scattering process in QCD at all orders in coupling constant.

Consider for example the $2 \rightarrow n$ partonic scattering process
\bea
k_1+k_2 \rightarrow k'_1+k'_2+...+k'_n
\label{gs}
\eea
in QCD where $k_1,k_2$ are the four-momenta of incoming gluons and $k'_1,k'_2,...,k'_n$ are the four-momenta of outgoing gluons. The initial state $|i>$ and the final state $f>$ for the above scattering process are given by
\bea
|i>=|k_1,k_2>,~~~~~~~~~~|f>=|k'_1,k'_2,...,k'_n>.
\label{ifi}
\eea
In this paper we will neglect the quarks but the inclusion of quarks is straightforward.

By using the LSZ reduction formula, the S-matrix element for the partonic scattering process in eq. (\ref{gs}) at all orders in coupling constant in QCD is given by
\bea
&&<f|i> = [G(-k'_1)]^{-1} [G(-k'_2)]^{-1}...[G(-k'_n)]^{-1}[G(k_2)]^{-1}[G(k_1)]^{-1} G(-k'_1,-k'_2,...,-k'_n,k_1,k_2)\nonumber \\
\label{hh33}
\eea
where $G(k)$ is the renormalized (full) propagator of gluon in momentum space and $G(k_1,...,k_n)$ is the renormalized n-point connected green's function of gluon in the momentum space. In eq. (\ref{hh33}) [and throughout this paper] the suppression of color and Lorentz indices is understood.
For simplicity we have included the finite factors such as the relevant sum over polarization vectors and color factors in the partonic cross section
\bea
{\hat \sigma} \propto |<f|i>|^2
\label{cr}
\eea
instead of the S-matrix element in eq. (\ref{hh33}) so that the S-matrix element in eq. (\ref{hh33}) is expressed in terms of the green's functions only.

In perturbative quantum chromodynamics (pQCD) the ultra violet (UV) divergence occurs in the calculation of loop diagram when the momentum integration limit goes to infinity. The renormalization program is introduced to handle the UV divergence in pQCD \cite{ht,gr}. One finds that the one particle irreducible (1PI) diagram is the basic building block in pQCD where the UV divergence occurs when the loop momentum integration limit goes to infinity. Hence as far as renormalization is concerned it is sufficient to study the UV divergence of the the N-point one particle irreducible (1PI) proper vertex function $\Gamma[k_1,...,k_N]$ in QCD.

From this point of view, as far as the renormalization of the S-matrix element is concerned, it is useful to express the S-matrix element in eq. (\ref{hh33}) in terms of the N-point 1PI proper vertex functions ${\bar \Gamma}[k_1,...,k_N]$ instead of the n-point connected green's function $G(k_1,...,k_n)$ at all orders in coupling constant where $N\le n$.

In coordinate space the eq. (\ref{hh33}) can be written as
\bea
&&<f|i> =\int d^4x'_1... \int d^4x'_n \int d^4x_2  \int d^4x_1 ~e^{i k'_1 \cdot x'_1+...+i k'_n \cdot x'_n-i k_2 \cdot x_2-i k_1 \cdot x_1} \int d^4y'_1...\int d^4y'_n \int d^4y_2 \int d^4y_1 \nonumber \\
&& \times [G(x'_1,y'_1)]^{-1}... [G(x'_n,y'_n)]^{-1}[G(x_2,y_2)]^{-1}[G(x_1,y_1)]^{-1}~G(y'_1,...,y'_n,y_2,y_1)
\label{nbg}
\eea
where $G(x_1,x_2)$ is the renormalized (full) propagator of gluon in coordinate space and $G(x_1,...,x_n)$ is the renormalized n-point connected green's function of gluon in coordinate space. Note that eq. (\ref{nbg}) is suitable to study factorization of infrared (IR) and collinear divergences in QCD at all orders in coupling constant \cite{n,s,gs}.

It can be mentioned here that the 2-point 1PI vertex function  $\Gamma[x_1,x_2]$ is the inverse of the (full) propagator (the 2-point connected Green's function $G(x_1,x_2)$) and the 3-point connected green's function $G(x_1,x_2,x_3)$ is expressed in terms of the 3-point 1PI vertex function $\Gamma[x_1,x_2,x_3]$ by adding (full) propagators to the external legs \cite{ab1}. Similarly, the 4-point connected green's function $G(x_1,x_2,x_3,x_4)$ is expressed in terms of the 4-point 1PI vertex function $\Gamma[x_1,x_2,x_3,x_4]$ and the 3-point 1PI vertex function $\Gamma[x_1,x_2,x_3]$ and the (full) propagator $G(x_1,x_2)$ \cite{ab1}.

In this paper we express the 5-point connected green's function $G(x_1,x_2,x_3,x_4,x_5)$ of gluon in terms of the 5-point 1PI vertex function $\Gamma[x_1,x_2,x_3,x_4,x_5]$ and the 4-point 1PI vertex function $\Gamma[x_1,x_2,x_3,x_4]$ and the 3-point 1PI vertex function $\Gamma[x_1,x_2,x_3]$ and the (full) propagator $G(x_1,x_2)$ at all orders in coupling constant by using the path integral formulation of QCD.

We also perform our calculation in the momentum space and express the 5-point connected green's function $G(k_1,k_2,k_3,k_4,k_5)$ of gluon in terms of the 5-point 1PI proper vertex function ${\bar \Gamma}[k_1,k_2,k_3,k_4,k_5]$ and the 4-point 1PI proper vertex function ${\bar \Gamma}[k_1,k_2,k_3,k_4]$ and the 3-point 1PI proper vertex function ${\bar \Gamma}[k_1,k_2,k_3]$ and the (full) propagator $G(k)$ at all orders in coupling constant by using the path integral formulation of QCD, see eq. (\ref{5gk}). We use this in the LSZ reduction formula and express the S-matrix element for the $gg \rightarrow ggg$ scattering process at all orders in coupling constant in QCD in terms of 5-point, 4-point, 3-point 1PI proper vertex functions and the (full) propagator.

We will provide a derivation of eq. (\ref{5gk}) in this paper.

The paper is organized as follows. In section II we describe the generating functional $Z[J,\eta,{\bar \eta}]$ in QCD in the path integral formulation. In section III we obtain the n-point connected green's function $G(x_1,...,x_n)$ and the n-point 1PI vertex function ${ \Gamma}[x_1,...,x_n]$ of gluon from the generating functional in QCD by using the the path integral formulation. In section IV we express 5-point connected Green's function of gluon in terms of 5-point, 4-point, 3-point 1PI vertex functions and the (full) propagator in coordinate space at all orders in coupling constant in QCD. In section V we express 5-point connected Green's function of gluon in terms of 5-point, 4-point, 3-point 1PI proper vertex functions and the (full) propagator in momentum space at all orders in coupling constant in QCD. In section VI we use this in the LSZ reduction formula and express the S-matrix element for the $gg \rightarrow ggg$ scattering process at all orders in coupling constant in QCD in terms of 5-point, 4-point, 3-point 1PI proper vertex functions and the (full) propagator. Section VII contains conclusion.

\section{ Generating Functional in the Path Integral Formulation of QCD  }

We denote the gluon field by $Q^{\mu a}(x)$ where $\mu=0,1,2,3$ is the Lorentz index and $a=1,...,8$ is the color index. In the path integral formulation the generating functional $Z[J,\eta,{\bar \eta}]$ in QCD is given by \cite{ab}
\bea
&&Z[J,\eta,{\bar \eta}]=\int [d{\bar \psi}][d\psi] [dQ] ~{\rm det}(\frac{\delta \partial^\mu Q_\mu^d}{\delta \omega^e})\nonumber \\
&&\times e^{i\int d^4x [-\frac{1}{4}F_{\mu \nu }^{d^2}[Q] - \frac{1}{2\alpha} (\partial_\mu Q^{\mu d}(x))^2 +{\bar \psi}(x)[i\gamma^\mu \partial_\mu +gT^d \gamma^\mu Q_\mu^d(x) -m]\psi(x) + {\bar \psi}(x) \cdot \eta(x) +{\bar \eta}(x) \cdot \psi(x)+ J(x) \cdot Q(x)]}\nonumber \\
\label{q1v}
\eea
where $\alpha$ is the gauge fixing parameter and
\bea
&&F_{\mu \nu }^{d^2}[Q] =[\partial_\mu Q_\nu^d(x) -  \partial_\nu Q_\mu^d(x) + gf^{dbc} Q_\mu^b(x) Q_\nu^c(x)] \nonumber \\
&&\times [\partial^\mu Q^{\nu d}(x) -  \partial^\nu Q^{\mu d}(x) + gf^{dae} Q^{\mu a}(x) Q^{\nu e}(x)].
\label{q3v}
\eea
In eq. (\ref{q1v}) the determinant ${\rm det}( \frac{\delta \partial^\mu Q_\mu^d}{\delta \omega^e})$ can be expressed in terms of the path integration over the ghost fields but we will directly work with ${\rm det}(\frac{\delta \partial^\mu Q_\mu^d}{\delta \omega^e})$ in eq. (\ref{q1v}). The external source to the quark field $\psi_i(x)$ is ${\bar \eta}_i(x)$ and the external source to the gluon field $Q^{\mu a}(x)$ is $J^{\mu a}(x)$.

The n-point connected green's function $G(x_1,...,x_n)$ and the n-point 1PI vertex function $\Gamma[x_1,...,x_n]$ of gluon at all orders in coupling constant in QCD can be generated from the generating functional in eq. (\ref{q1v}) by using the path integral formulation of QCD.

\section{N-point connected green's function of gluon and the n-point 1PI vertex function of gluon in QCD}

As mentioned above the n-point green's function $G(x_1,...,x_n)$ of gluon at all orders in coupling constant in QCD can be obtained from the generating functional in eq. (\ref{q1v}) by using the path integral formulation of QCD. From eq. (\ref{q1v}) we find that the n-point connected Green's function $G(x_1,...,x_n)$ of gluon at all orders in coupling constant in QCD is given by
\bea
G(x_1,...,x_n)=\frac{1}{i^{n-1}}~ \frac{ \delta^n W[J,\eta,{\bar \eta}]}{\delta J(x_1)...\delta J(x_n)}|_{\eta ={\bar \eta} =J=0}
\label{w1v}
\eea
where $W[J,\eta,{\bar \eta}]$ is related to $Z[J,\eta,{\bar \eta}]$ in eq. (\ref{q1v}) via the equation
\bea
W[J,\eta,{\bar \eta}=-i~{\rm ln}Z[J,\eta,{\bar \eta}].
\label{wzv}
\eea
The n-point Green's function of gluon in QCD obeys the invariance
\bea
G(x_1,x_2,...,x_n) = G(x_1+y,x_2+y,...,x_n+y).
\label{giv}
\eea

From eq. (\ref{wzv}) we find that the effective action functional in QCD is given by
\bea
\Gamma[<Q>,<\eta>,<{\bar \eta}>]=W[J,\eta,{\bar \eta}]-\int d^4x[J(x)\cdot <Q(x)>+{\bar \eta}(x) \cdot <\psi(x)>+<{\bar \psi}(x)> \cdot \eta(x) ] \nonumber \\
\label{w4v}
\eea
which generates the n-point 1PI vertex function $\Gamma[x_1,...,x_n]$ of gluon at all orders in coupling constant in QCD given by
\bea
\Gamma[x_1,...,x_n]=\frac{1}{i^{n-1}} \frac{\delta^n \Gamma[<Q>,<\eta>,<{\bar \eta}>]}{\delta <Q(x_1)>...\delta <Q(x_n)>}|_{<Q>=<\eta>=<{\bar \eta}>=0}
\label{w5v}
\eea
where
\bea
<Q(x)>= \frac{\delta W[J,\eta,{\bar \eta}]}{\delta J(x)},~~~~~<\psi(x)>= \frac{\delta W[J,\eta,{\bar \eta}]}{\delta {\bar \eta}(x)},~~~~~<{\bar \psi}(x)>= \frac{\delta W[J,\eta,{\bar \eta}]}{\delta { \eta}(x)}.
\label{w3v}
\eea

\section{5-point connected green's function and the 5-point, 4-point and 3-point 1PI vertex functions of Gluon in coordinate space }

As mentioned in the introduction the 2-point 1PI vertex function  $\Gamma[x_1,x_2]$ is the inverse of the (full) propagator (the 2-point connected Green's function $G(x_1,x_2)$) and the 3-point connected green's function $G(x_1,x_2,x_3)$ is expressed in terms of the 3-point 1PI vertex function $\Gamma[x_1,x_2,x_3]$ by adding (full) propagators to the external legs in \cite{ab1} which can be easily verified from eqs. (\ref{w1v}) and (\ref{w5v}). Similarly, the 4-point connected green's function $G(x_1,x_2,x_3,x_4)$ is expressed in terms of the 4-point 1PI vertex function $\Gamma[x_1,x_2,x_3,x_4]$ and the 3-point 1PI vertex function $\Gamma[x_1,x_2,x_3]$ and the (full) propagator $G(x_1,x_2)$ in \cite{ab1} which can be verified from eqs. (\ref{w1v}) and (\ref{w5v}).

In this section we will express the 5-point connected green's function $G(x_1,x_2,x_3,x_4,x_5)$ of gluon in terms of the 5-point 1PI vertex function $\Gamma[x_1,x_2,x_3,x_4,x_5]$ and the 4-point 1PI vertex function $\Gamma[x_1,x_2,x_3,x_4]$ and the 3-point 1PI vertex function $\Gamma[x_1,x_2,x_3]$ and the (full) propagator $G(x_1,x_2)$ at all orders in coupling constant in QCD.

As discussed in the previous section, in the path integral formulation of QCD, the n-point connected Green's function $G(x_1,...,x_n)$ of gluon at all orders in coupling constant in QCD is given by eq. (\ref{w1v}) and the n-point 1PI vertex function $\Gamma[x_1,...,x_n]$ of gluon at all orders in coupling constant in QCD is given by eq. (\ref{w5v}). Hence after doing a lengthy but straightforward calculation we find from eqs. (\ref{w1v}) and (\ref{w5v}) that
\bea
&&G(x_1,x_2,x_3,x_4,x_5)= \nonumber \\
&&\int d^4x'_1 \int d^4x'_2 \int d^4x'_3 \int d^4x'_4 \int d^4x'_5 G(x_1,x'_1)G(x_2,x'_2)G(x_3,x'_3) G(x_4,x'_4) G(x_5,x'_5) \Gamma[x'_1,x'_2,x'_3,x'_4,x'_5]\nonumber \\
&&  +\int d^4x \int d^4y \int d^4z \int d^4w \int d^4w_1 \int d^4w_2 \int d^4w_3 [\nonumber \\
&&G(x_1,w_1)G(x_5,w_2)G(x,w_3)\Gamma[w_1,w_2,w_3]G(x_2,y)G(x_3,z)G(x_4,w)\Gamma[x,y,z,w]\nonumber \\
&&+G(x_1,w_1)G(x_2,w_2)G(y,w_3)\Gamma[w_1,w_2,w_3]G(x_5,x)G(x_3,z)G(x_4,w)\Gamma[x,y,z,w]\nonumber \\
&&+G(x_1,w_1)G(x_3,w_2)G(z,w_3)\Gamma[w_1,w_2,w_3]G(x_5,x) G(x_2,y)G(x_4,w)\Gamma[x,y,z,w] \nonumber \\
&&+G(x_1,w_1)G(x_4,w_2)G(w,w_3)\Gamma[w_1,w_2,w_3]G(x_5,x) G(x_2,y)G(x_3,z)\Gamma[x,y,z,w] \nonumber \\
&&+G(x_1,x) G(x_5,y)G(x_2,z)\Gamma[x,y,z,w]G(w,w_1)G(x_3,w_2)G(x_4,w_3)\Gamma[w_1,w_2,w_3]\nonumber \\
&&+G(x_5,w_1)G(x_2,w_2)G(y,w_3)\Gamma[w_1,w_2,w_3]G(x_1,x) G(x_3,z)G(x_4,w)\Gamma[x,y,z,w]\nonumber \\
&& +G(x_1,x) G(x_5,y)G(x_3,z)\Gamma[x,y,z,w]G(x_2,w_1)G(w,w_2)G(x_4,w_3)\Gamma[w_1,w_2,w_3]\nonumber \\
&&+G(x_5,w_1)G(x_3,w_2)G(z,w_3)\Gamma[w_1,w_2,w_3]G(x_1,x) G(x_2,y)G(x_4,w)\Gamma[x,y,z,w] \nonumber \\
&& +G(x_1,x) G(x_5,y)G(x_4,z)\Gamma[x,y,z,w]G(x_2,w_1)G(x_3,w_2)G(w,w_3)\Gamma[w_1,w_2,w_3] \nonumber \\
&&+G(x_5,w_1)G(x_4,w_2)G(w,w_3)\Gamma[w_1,w_2,w_3]G(x_1,x) G(x_2,y)G(x_3,z)\Gamma[x,y,z,w]] \nonumber \\
&&  +\int d^4x \int d^4y \int d^4z \int d^4w \int d^4w_1 \int d^4w_2 \int d^4w_3 \int d^4y' \int d^4z' [ \nonumber \\
&&G(x_1,w_1)G(x_5,w_2)G(x,w_3)\Gamma[w_1,w_2,w_3] G(x_2,y)G(x_3,z)G(x_4,w)\Gamma[x,y,y']G(y',z')\Gamma[z',z,w]\nonumber \\
&&+G(x_1,w_1)G(x_2,w_2)G(y,w_3)\Gamma[w_1,w_2,w_3]G(x_5,x) G(x_3,z)G(x_4,w)\Gamma[x,y,y']G(y',z')\Gamma[z',z,w]\nonumber \\
&&+G(x_1,w_1)G(x_3,w_2)G(z,w_3)\Gamma[w_1,w_2,w_3]G(x_5,x) G(x_2,y)G(x_4,w)\Gamma[x,y,y']G(y',z')\Gamma[z',z,w]\nonumber \\
&&+G(x_1,w_1)G(x_4,w_2)G(w,w_3)\Gamma[w_1,w_2,w_3] G(x_5,x) G(x_2,y)G(x_3,z)\Gamma[x,y,y']G(y',z')\Gamma[z',z,w]\nonumber \\
&&+G(x_1,x) G(x_5,y)G(x_2,z)G(w_1,w)\Gamma[x,y,y']G(y',z')\Gamma[z',z,w]G(x_3,w_2)G(x_4,w_3)\Gamma[w_1,w_2,w_3]\nonumber \\
&&+G(x_5,w_1)G(x_2,w_2)G(y,w_3)\Gamma[w_1,w_2,w_3]G(x_1,x)G(x_3,z)G(x_4,w)\Gamma[x,y,y']G(y',z')\Gamma[z',z,w]\nonumber \\
&& +G(x_1,x) G(x_5,y)G(x_3,z)G(w_2,w)\Gamma[x,y,y']G(y',z')\Gamma[z',z,w]G(x_2,w_1)G(x_4,w_3)\Gamma[w_1,w_2,w_3]\nonumber \\
&&+G(x_5,w_1)G(x_3,w_2)G(z,w_3)\Gamma[w_1,w_2,w_3]G(x_1,x) G(x_2,y)G(x_4,w)\Gamma[x,y,y']G(y',z')\Gamma[z',z,w]\nonumber \\
&&+G(x_1,x)G(x_5,y)G(x_4,z)G(w_3,w)\Gamma[x,y,y']G(y',z')\Gamma[z',z,w]G(x_2,w_1)G(x_3,w_2)\Gamma[w_1,w_2,w_3] \nonumber \\
&&+G(x_5,w_1)G(x_4,w_2)G(w,w_3)\Gamma[w_1,w_2,w_3]G(x_1,x) G(x_2,y)G(x_3,z)\Gamma[x,y,y']G(y',z')\Gamma[z',z,w] \nonumber \\
&&+G(x_1,w_1)G(x_5,w_2)\Gamma[w_1,w_2,w_3]G(w_3,x) G(x_2,y)G(x_3,z)G(x_4,w)\Gamma[x,z,y']G(y',z')\Gamma[y,z',w]\nonumber \\
&&+G(x_1,w_1)G(x_2,w_2)G(y,w_3)\Gamma[w_1,w_2,w_3]G(x_5,x)G(x_3,z)G(x_4,w)\Gamma[x,z,y']G(y',z')\Gamma[y,z',w]\nonumber \\
&&+G(x_1,w_1)G(x_3,w_2)\Gamma[w_1,w_2,w_3]G(x_5,x) G(x_2,y)G(w_3,z)G(x_4,w)\Gamma[x,z,y']G(y',z')\Gamma[y,z',w]\nonumber \\
&&+G(x_1,w_1)G(x_4,w_2)\Gamma[w_1,w_2,w_3]G(x_5,x) G(x_2,y)G(x_3,z)G(w_3,w)\Gamma[x,z,y']G(y',z')\Gamma[y,z',w]\nonumber \\
&&+G(x_1,x) G(x_5,y)G(x_2,z)G(w_1,w)\Gamma[x,z,y']G(y',z')\Gamma[y,z',w]G(x_3,w_2)G(x_4,w_3)\Gamma[w_1,w_2,w_3]\nonumber \\
&&+G(x_5,w_1)G(x_2,w_2)\Gamma[w_1,w_2,w_3]G(x_1,x) G(w_3,y)G(x_3,z)G(x_4,w)\Gamma[x,z,y']G(y',z')\Gamma[y,z',w]\nonumber \\
&& +G(x_1,x) G(x_5,y)G(x_3,z)\Gamma[x,z,y']G(y',z')\Gamma[y,z',w]G(x_2,w_1)G(w,w_2)G(x_4,w_3)\Gamma[w_1,w_2,w_3]\nonumber \\
&&+G(x_5,w_1)G(x_3,w_2)G(z,w_3)\Gamma[w_1,w_2,w_3]G(x_1,x) G(x_2,y)G(x_4,w)\Gamma[x,z,y']G(y',z')\Gamma[y,z',w]\nonumber \\
&&+G(x_1,x) G(x_5,y)G(x_4,z)\Gamma[x,z,y']G(y',z')\Gamma[y,z',w]G(x_2,w_1)G(x_3,w_2)G(w,w_3)\Gamma[w_1,w_2,w_3] \nonumber \\
&&+G(x_5,w_1)G(x_4,w_2)\Gamma[w_1,w_2,w_3]G(x_1,x) G(x_2,y)G(x_3,z)G(w_3,w)\Gamma[x,z,y']G(y',z')\Gamma[y,z',w]\nonumber \\
&&+G(x_1,w_1)G(x_5,w_2)G(x,w_3)\Gamma[w_1,w_2,w_3]G(x_2,y)G(x_3,z)G(x_4,w)\Gamma[x,w,y']G(y',z')\Gamma[y,z,z']\nonumber \\
&&+G(x_1,w_1)G(x_2,w_2)G(y,w_3)\Gamma[w_1,w_2,w_3]G(x_5,x)G(x_3,z)G(x_4,w)\Gamma[x,w,y']G(y',z')\Gamma[y,z,z']\nonumber \\
&&+G(x_1,w_1)G(x_3,w_2)\Gamma[w_1,w_2,w_3]G(x_5,x) G(x_2,y)G(w_3,z)G(x_4,w)\Gamma[x,w,y']G(y',z')\Gamma[y,z,z']\nonumber \\
&&+G(x_1,w_1)G(x_4,w_2)\Gamma[w_1,w_2,w_3]G(x_5,x) G(x_2,y)G(x_3,z)G(w_3,w)\Gamma[x,w,y']G(y',z')\Gamma[y,z,z']\nonumber \\
&&+G(x_1,x) G(x_5,y)G(x_2,z)\Gamma[x,w,y']G(y',z')\Gamma[y,z,z']G(w,w_1)G(x_3,w_2)G(x_4,w_3)\Gamma[w_1,w_2,w_3]\nonumber \\
&&+G(x_5,w_1)G(x_2,w_2)\Gamma[w_1,w_2,w_3]G(x_1,x) G(w_3,y)G(x_3,z)G(x_4,w)\Gamma[x,w,y']G(y',z')\Gamma[y,z,z']\nonumber \\
&& +G(x_1,x) G(x_5,y)G(x_3,z)\Gamma[x,w,y']G(y',z')\Gamma[y,z,z']G(x_2,w_1)G(w,w_2)G(x_4,w_3)\Gamma[w_1,w_2,w_3]\nonumber \\
&&+G(x_5,w_1)G(x_3,w_2)G(z,w_3)\Gamma[w_1,w_2,w_3]G(x_1,x) G(x_2,y)G(x_4,w)\Gamma[x,w,y']G(y',z')\Gamma[y,z,z']\nonumber \\
&& +G(x_1,x) G(x_5,y)G(x_4,z)\Gamma[x,w,y']G(y',z')\Gamma[y,z,z']G(x_2,w_1)G(x_3,w_2)G(w,w_3)\Gamma[w_1,w_2,w_3]\nonumber \\
&&+G(x_5,w_1)G(x_4,w_2)G(w,w_3)\Gamma[w_1,w_2,w_3]G(x_1,x) G(x_2,y)G(x_3,z)\Gamma[x,w,y']G(y',z')\Gamma[y,z,z']\nonumber \\
&&+G(x_1,w_1)G(x_5,w_1)\Gamma[w_1,w_2,w_3]G(w_2,x)G(x_2,y)\Gamma[x,y,z]G(z,w)G(x_3,y')G(x_4,z')\Gamma[w,y',z'] \nonumber \\
&&+G(x_1,w_1)G(x_5,w_2)\Gamma[w_1,w_2,w_3]G(w_3,x)G(x_3,y)\Gamma[x,y,z]G(x_2,w)G(z,y')G(x_4,z')\Gamma[w,y',z']\nonumber \\
&&+G(x_1,w_1)G(x_5,w_2)\Gamma[w_1,w_2,w_3]G(w_3,x)G(x_4,y)\Gamma[x,y,z]G(x_2,w)G(x_3,y')G(z,z')\Gamma[w,y',z']\nonumber \\
&&+G(x_1,w_1)G(x_2,w_2)\Gamma[w_1,w_2,w_3]G(x_5,x)G(w_3,y)\Gamma[x,y,z]G(z,w)G(x_3,y')G(x_4,z')\Gamma[w,y',z']\nonumber \\
&&+G(x_1,w_1)\Gamma[w_1,w_2,w_3]G(x_5,x)G(x_2,y)G(w_3,z)\Gamma[x,y,z]G(w_2,w)G(x_3,y')G(x_4,z')\Gamma[w,y',z']\nonumber \\
&& +G(x_5,w_1)G(x_2,w_2)\Gamma[w_1,w_2,w_3]G(x_1,x)G(x_3,y)\Gamma[x,y,z]G(w_3,w)G(z,y')G(x_4,z')\Gamma[w,y',z']\nonumber \\
&& +G(x_5,w_1)G(x_2,w_2)\Gamma[w_1,w_2,w_3]G(x_1,x)G(x_4,y)\Gamma[x,y,z]G(w_3,w)G(x_3,y')G(z,z')\Gamma[w,y',z']\nonumber \\
&&+G(x_1,w_1)G(x_2,w_2)\Gamma[w_1,w_2,w_3]G(x_5,x)G(x_3,y)\Gamma[x,y,z]G(w_3,w)G(z,y')G(x_4,z')\Gamma[w,y',z']\nonumber \\
&& +G(x_1,w_1)\Gamma[w_1,w_2,w_3]G(x_5,x)G(x_3,y)G(w_3,z)\Gamma[x,y,z]G(x_2,w)G(w_2,y')G(x_4,z')\Gamma[w,y',z']\nonumber \\
&& + G(x_5,w_1)G(z,w_2)\Gamma[w_1,w_2,w_3]G(x_1,x)G(x_3,y)\Gamma[x,y,z]G(x_2,w)G(w_3,y')G(x_4,z')\Gamma[w,y',z']\nonumber \\
&& +G(x_5,w_1)G(x_3,w_2)\Gamma[w_1,w_2,w_3]G(x_1,x)G(x_4,y)\Gamma[x,y,z]G(x_2,w)G(w_3,y')G(z,z')\Gamma[w,y',z']\nonumber \\
&&+G(x_1,w_1)G(x_2,w_2)\Gamma[w_1,w_2,w_3]G(x_5,x)G(x_4,y)\Gamma[x,y,z]G(w_3,w)G(x_3,y')G(z,z')\Gamma[w,y',z']\nonumber \\
&& +G(x_1,w_1)\Gamma[w_1,w_2,w_3]G(x_5,x)G(x_4,y)G(w_3,z)\Gamma[x,y,z]G(x_2,w)G(x_3,y')G(w_2,z')\Gamma[w,y',z']\nonumber \\
&& +G(x_5,w_1)G(x_4,w_2)\Gamma[w_1,w_2,w_3]G(x_1,x)G(x_3,y)\Gamma[x,y,z]G(x_2,w)G(z,y')G(w_3,z')\Gamma[w,y',z']\nonumber \\
&& +G(x_5,w_1)G(z,w_2)\Gamma[w_1,w_2,w_3]G(x_1,x)G(x_4,y)\Gamma[x,y,z]G(x_2,w)G(x_3,y')G(w_3,z')\Gamma[w,y',z'] ]\nonumber \\
\label{5gx}
\eea
which is the expression of the 5-point connected green's function $G(x_1,x_2,x_3,x_4,x_5)$ of gluon in coordinate space in terms of the 5-point 1PI vertex function $\Gamma[x_1,x_2,x_3,x_4,x_5]$ and the 4-point 1PI vertex function $\Gamma[x_1,x_2,x_3,x_4]$ and the 3-point 1PI vertex function $\Gamma[x_1,x_2,x_3]$ and the (full) propagator $G(x_1,x_2)$ at all orders in coupling constant in QCD.

\section{5-point connected green's function and the 5-point, 4-point and 3-point 1PI Proper vertex functions of Gluon in momentum space}

In this section we perform the calculation in the momentum space and express the 5-point connected green's function $G(k_1,k_2,k_3,k_4,k_5)$ of gluon in terms of the 5-point 1PI proper vertex function ${\bar \Gamma}[k_1,k_2,k_3,k_4,k_5]$ and the 4-point 1PI proper vertex function ${\bar \Gamma}[k_1,k_2,k_3,k_4]$ and the 3-point 1PI proper vertex function ${\bar \Gamma}[k_1,k_2,k_3]$ and the (full) propagator $G(k)$ at all orders in coupling constant in QCD.

Note that since the sum of total momentum is zero in the 1PI vertex function, the n-point 1PI vertex function $\Gamma[k_1,k_2,...,k_n]$ in momentum space is related to the n-point 1PI proper vertex function ${\bar \Gamma}[k_1,k_2,...,k_n]$ via the relation
\bea
\Gamma[k_1,k_2,...,k_n]=\delta^{(4)}(k_1+k_2+...+k_n)~{\bar \Gamma}[k_1,k_2,...,k_n].
\label{pgv}
\eea

Hence in the path integral formulation of QCD we find from eqs.  (\ref{giv}), (\ref{pgv}) and (\ref{5gx}) that
\bea
&&[G(k_1)]^{-1} [G(k_2)]^{-1}[G(k_3)]^{-1}[G(k_4)]^{-1}[G(k_5)]^{-1}G(k_1,k_2,k_3,k_4,k_5)= \delta^{(4)}(k_1+k_2+k_3+k_4+k_5)\nonumber \\
&&\times [{\bar \Gamma}[k_1,k_2,k_3,k_4,k_5]~\nonumber \\
&&  +{\bar \Gamma}[k_1,k_5,-k_1-k_5]G(-k_1-k_5){\bar \Gamma}[k_1+k_5,k_2,k_3,k_4]\nonumber \\
&& +{\bar \Gamma}[k_1,k_2,-k_1-k_2]G(-k_1-k_2){\bar \Gamma}[k_5,k_1+k_2,k_3,k_4]\nonumber \\
&& +{\bar \Gamma}[k_1,k_3,-k_1-k_3]G(-k_1-k_3){\bar \Gamma}[k_5,k_2,k_1+k_3,k_4]\nonumber \\
&& +{\bar \Gamma}[k_1,k_4,-k_1-k_4]G(-k_1-k_4){\bar \Gamma}[k_5,k_2,k_3,k_1+k_4]\nonumber \\
&& +{\bar \Gamma}[-k_3-k_4,k_3,k_4]G(-k_3-k_4){\bar \Gamma}[k_1,k_5,k_2,k_3+k_4]\nonumber \\
&& +{\bar \Gamma}[k_5,k_2,-k_5-k_2]G(-k_5-k_2){\bar \Gamma}[k_1,k_5+k_2,k_3,k_4]\nonumber \\
&& +{\bar \Gamma}[k_2,-k_2-k_4,k_4]G(-k_2-k_4){\bar \Gamma}[k_1,k_5,k_3,k_2+k_4]\nonumber \\
&& +{\bar \Gamma}[k_5,k_3,-k_5-k_3]G(-k_5-k_3){\bar \Gamma}[k_1,k_2,k_5+k_3,k_4]\nonumber \\
&& +{\bar \Gamma}[k_2,k_3,-k_2-k_3]G(-k_2-k_3){\bar \Gamma}[k_1,k_5,k_4,k_2+k_3]\nonumber \\
&& +{\bar \Gamma}[k_5,k_4,-k_5-k_4]G(-k_5-k_4){\bar \Gamma}[k_1,k_2,k_3,k_5+k_4]\nonumber \\
&& +{\bar \Gamma}[k_1,k_5,-k_1-k_5]G(-k_1-k_5){\bar \Gamma}[k_1+k_5,k_2,k_3+k_4]G(-k_3-k_4){\bar \Gamma}[-k_3-k_4,k_3,k_4]\nonumber \\
&& +{\bar \Gamma}[k_1,k_2,-k_1-k_2]G(-k_1-k_2){\bar \Gamma}[k_5,k_1+k_2,k_3+k_4]G(-k_3-k_4){\bar \Gamma}[-k_3-k_4,k_3,k_4]\nonumber \\
&& + {\bar \Gamma}[k_1,k_3,-k_1-k_3]G(-k_1-k_3){\bar \Gamma}[k_5,k_2,-k_5-k_2]G(k_5+k_2){\bar \Gamma}[k_5+k_2,k_1+k_3,k_4]\nonumber \\
&& +{\bar \Gamma}[k_1,k_4,-k_1-k_4]G(-k_1-k_4){\bar \Gamma}[k_5,k_2,-k_5-k_2]G(k_5+k_2){\bar \Gamma}[k_5+k_2,k_3,k_1+k_4]\nonumber \\
&& + {\bar \Gamma}[-k_3-k_4,k_3,k_4]G(k_3+k_4){\bar \Gamma}[k_1,k_5,-k_1-k_5]G(k_1+k_5){\bar \Gamma}[k_1+k_5,k_2,k_3+k_4]\nonumber \\
&& +{\bar \Gamma}[k_5,k_2,-k_5-k_2]G(-k_5-k_2){\bar \Gamma}[k_1,k_5+k_2,k_3+k_4]G(-k_3-k_4){\bar \Gamma}[-k_3-k_4,k_3,k_4]\nonumber \\
&& +{\bar \Gamma}[k_2,-k_2-k_4,k_4]G(k_2+k_4){\bar \Gamma}[k_1,k_5,-k_1-k_5]G(k_1+k_5){\bar \Gamma}[k_1+k_5,k_3,k_2+k_4]\nonumber \\
&& +{\bar \Gamma}[k_5,k_3,-k_5-k_3]G(-k_5-k_3){\bar \Gamma}[k_1,k_2,-k_1-k_2]G(k_1+k_2){\bar \Gamma}[k_1+k_2,k_5+k_3,k_4]\nonumber \\
&& +{\bar \Gamma}[k_2,k_3,-k_2-k_3]G(k_2+k_3){\bar \Gamma}[k_1,k_5,-k_1-k_5]G(k_1+k_5){\bar \Gamma}[k_1+k_5,k_4,k_2+k_3]\nonumber \\
&& +{\bar \Gamma}[k_5,k_4,-k_5-k_4]G(-k_5-k_4){\bar \Gamma}[k_1,k_2,-k_1-k_2]G(k_1+k_2){\bar \Gamma}[k_1+k_2,k_3,k_5+k_4]\nonumber \\
&& +{\bar \Gamma}[k_1,k_5,-k_1-k_5]G(k_1+k_5){\bar \Gamma}[k_1+k_5,k_3,k_2+k_4]G(-k_2-k_4){\bar \Gamma}[k_2,-k_2-k_4,k_4]\nonumber \\
&& +{\bar \Gamma}[k_1,k_2,-k_1-k_2]G(-k_1-k_2){\bar \Gamma}[k_5,k_3,-k_5-k_3]G(k_5+k_3){\bar \Gamma}[k_1+k_2,k_5+k_3,k_4]\nonumber \\
&& +{\bar \Gamma}[k_1,k_3,-k_1-k_3]G(k_1+k_3){\bar \Gamma}[k_5,k_1+k_3,k_2+k_4]G(-k_2-k_4){\bar \Gamma}[k_2,-k_2-k_4,k_4]\nonumber \\
&& +{\bar \Gamma}[k_1,k_4,-k_1-k_4]G(k_1+k_4){\bar \Gamma}[k_5,k_3,-k_5-k_3]G(k_5+k_3){\bar \Gamma}[k_2,k_5+k_3,k_1+k_4]\nonumber \\
&& +{\bar \Gamma}[-k_3-k_4,k_3,k_4]G(k_3+k_4){\bar \Gamma}[k_1,k_2,-k_1-k_2]G(k_1+k_2){\bar \Gamma}[k_5,k_1+k_2,k_3+k_4]\nonumber \\
&& +{\bar \Gamma}[k_5,k_2,-k_5-k_2]G(k_5+k_2){\bar \Gamma}[k_1,k_3,-k_1-k_3]G(k_1+k_3){\bar \Gamma}[k_5+k_2,k_1+k_3,k_4]\nonumber \\
&& +{\bar \Gamma}[k_2,-k_2-k_4,k_4]G(-k_2-k_4){\bar \Gamma}[k_1,k_3,-k_1-k_3]G(k_1+k_3){\bar \Gamma}[k_5,k_1+k_3,k_2+k_4]\nonumber \\
&& +{\bar \Gamma}[k_5,k_3,-k_5-k_3]G(-k_5-k_3){\bar \Gamma}[k_1,k_5+k_3,k_2+k_4]G(-k_2-k_4){\bar \Gamma}[k_2,-k_2-k_4,k_4]\nonumber \\
&& +{\bar \Gamma}[k_2,k_3,-k_2-k_3]G(-k_2-k_3){\bar \Gamma}[k_1,k_4,-k_1-k_4]G(-k_1-k_4){\bar \Gamma}[k_5,k_1+k_4,k_2+k_3]\nonumber \\
&& +{\bar \Gamma}[k_5,k_4,-k_5-k_4]G(k_5+k_4){\bar \Gamma}[k_1,k_3,-k_1-k_3]G(k_1+k_3){\bar \Gamma}[k_2,k_1+k_3,k_5+k_4]\nonumber \\
&& +{\bar \Gamma}[k_1,k_5,-k_1-k_5]G(-k_1-k_5){\bar \Gamma}[k_1+k_5,k_4,k_2+k_3]G(-k_2-k_3){\bar \Gamma}[k_2,k_3,-k_2-k_3]\nonumber \\
&& +{\bar \Gamma}[k_1,k_2,-k_1-k_2]G(-k_1-k_2){\bar \Gamma}[k_5,k_4,-k_5-k_4]G(k_5+k_4){\bar \Gamma}[k_1+k_2,k_3,k_5+k_4]\nonumber \\
&& +{\bar \Gamma}[k_1,k_3,-k_1-k_3]G(k_1+k_3){\bar \Gamma}[k_5,k_4,-k_5-k_4]G(-k_5-k_4){\bar \Gamma}[k_2,k_1+k_3,k_5+k_4]\nonumber \\
&& +{\bar \Gamma}[k_1,k_4,-k_1-k_4]G(k_1+k_4){\bar \Gamma}[k_5,k_1+k_4,k_2+k_3]G(-k_2-k_3){\bar \Gamma}[k_2,k_3,-k_2-k_3]\nonumber \\
&& +{\bar \Gamma}[-k_3-k_4,k_3,k_4]G(-k_3-k_4){\bar \Gamma}[k_1,k_3+k_4,k_5+k_2]G(-k_5-k_2){\bar \Gamma}[k_5,k_2,-k_5-k_2]\nonumber \\
&& +{\bar \Gamma}[k_5,k_2,-k_5-k_2]G(k_5+k_2){\bar \Gamma}[k_1,k_4,-k_1-k_4]G(k_1+k_4){\bar \Gamma}[k_5+k_2,k_3,k_1+k_4]\nonumber \\
&& +{\bar \Gamma}[k_2,-k_2-k_4,k_4]G(-k_2-k_4){\bar \Gamma}[k_1,k_2+k_4,k_5+k_3]G(-k_5-k_3){\bar \Gamma}[k_5,k_3,-k_5-k_3]\nonumber \\
&& +{\bar \Gamma}[k_5,k_3,-k_5-k_3]G(-k_5-k_3){\bar \Gamma}[k_1,k_4,-k_1-k_4]G(k_1+k_4){\bar \Gamma}[k_2,k_5+k_3,k_1+k_4]\nonumber \\
&& +{\bar \Gamma}[k_2,k_3,-k_2-k_3]G(-k_2-k_3){\bar \Gamma}[k_1,k_2+k_3,k_5+k_4]G(-k_5-k_4){\bar \Gamma}[k_5,k_4,-k_5-k_4]\nonumber \\
&& +{\bar \Gamma}[k_5,k_4,-k_5-k_4]G(-k_5-k_4){\bar \Gamma}[k_1,k_5+k_4,k_2+k_3]G(-k_2-k_3){\bar \Gamma}[k_2,k_3,-k_2-k_3]\nonumber \\
&& +{\bar \Gamma}[k_1,-k_1-k_5,k_5]G(k_1+k_5){\bar \Gamma}[k_1+k_5,k_2,k_3+k_4]G(-k_3-k_4){\bar \Gamma}[-k_3-k_4,k_3,k_4]\nonumber \\
&& +{\bar \Gamma}[k_1,k_5,-k_1-k_5]G(k_1+k_5){\bar \Gamma}[k_1+k_5,k_3,k_2+k_4]G(-k_2-k_4){\bar \Gamma}[k_2,-k_2-k_4,k_4]\nonumber \\
&& +{\bar \Gamma}[k_1,k_5,-k_1-k_5]G(k_1+k_5){\bar \Gamma}[k_1+k_5,k_4,k_2+k_3]G(-k_2-k_3){\bar \Gamma}[k_2,k_3,-k_2-k_3]\nonumber \\
&& +{\bar \Gamma}[k_1,k_2,-k_1-k_2]G(k_1+k_2){\bar \Gamma}[k_5,k_1+k_2,k_3+k_4]G(-k_3-k_4){\bar \Gamma}[-k_3-k_4,k_3,k_4]\nonumber \\
&& +{\bar \Gamma}[k_1,k_3+k_4,k_5+k_2]G(-k_5-k_2){\bar \Gamma}[k_5,k_2,-k_5-k_2]G(-k_3-k_4){\bar \Gamma}[-k_3-k_4,k_3,k_4]\nonumber \\
&& +{\bar \Gamma}[k_5,k_2,-k_5-k_2]G(k_5+k_2){\bar \Gamma}[k_1,k_3,-k_1-k_3]G(k_1+k_3){\bar \Gamma}[k_5+k_2,k_1+k_3,k_4]\nonumber \\
&& +{\bar \Gamma}[k_5,k_2,-k_5-k_2]G(k_5+k_2){\bar \Gamma}[k_1,k_4,-k_1-k_4]G(k_1+k_4){\bar \Gamma}[k_5+k_2,k_3,k_1+k_4]\nonumber \\
&& +{\bar \Gamma}[k_1,k_2,-k_1-k_2]G(k_1+k_2){\bar \Gamma}[k_5,k_3,-k_5-k_3]G(k_5+k_3){\bar \Gamma}[k_1+k_2,k_5+k_3,k_4]\nonumber \\
&& +{\bar \Gamma}[k_1,k_2+k_4,k_5+k_3]G(-k_5-k_3){\bar \Gamma}[k_5,k_3,-k_5-k_3]G(-k_2-k_4){\bar \Gamma}[k_2,-k_2-k_4,k_4]\nonumber \\
&& +{\bar \Gamma}[k_5,k_1+k_3,k_2+k_4]G(k_1+k_3){\bar \Gamma}[k_1,k_3,-k_1-k_3]G(-k_2-k_4){\bar \Gamma}[k_2,-k_2-k_4,k_4]\nonumber \\
&& +{\bar \Gamma}[k_5,k_3,-k_5-k_3]G(-k_5-k_3){\bar \Gamma}[k_1,k_4,-k_1-k_4]G(-k_1-k_4){\bar \Gamma}[k_2,k_5+k_3,k_1+k_4]\nonumber \\
&& +{\bar \Gamma}[k_1,k_2,-k_1-k_2]G(k_1+k_2){\bar \Gamma}[k_5,k_4,-k_5-k_4]G(k_5+k_4){\bar \Gamma}[k_1+k_2,k_3,k_5+k_4]\nonumber \\
&& +{\bar \Gamma}[k_1,k_2+k_3,k_5+k_4]G(-k_5-k_4){\bar \Gamma}[k_5,k_4,-k_5-k_4]G(-k_2-k_3){\bar \Gamma}[k_2,k_3,-k_2-k_3]\nonumber \\
&& +{\bar \Gamma}[k_5,k_4,-k_5-k_4]G(k_1+k_3){\bar \Gamma}[k_1,k_3,-k_1-k_3]G(k_5+k_4){\bar \Gamma}[k_2,k_1+k_3,k_5+k_4]\nonumber \\
&& + {\bar \Gamma}[k_5,k_1+k_4,k_2+k_3]G(k_1+k_4){\bar \Gamma}[k_1,k_4,-k_1-k_4]G(-k_2-k_3){\bar \Gamma}[k_2,k_3,-k_2-k_3]~]
\label{5gk}
\eea
which is the expression of the 5-point connected green's function $G(k_1,k_2,k_3,k_4,k_5)$ of gluon in momentum space in terms of the 5-point 1PI proper vertex function ${\bar \Gamma}[k_1,k_2,k_3,k_4,k_5]$ and the 4-point 1PI proper vertex function ${\bar \Gamma}[k_1,k_2,k_3,k_4]$ and the 3-point 1PI proper vertex function ${\bar \Gamma}[k_1,k_2,k_3]$ and the (full) propagator $G(k)$ at all orders in coupling constant in QCD.

\section{ 5-Point 1PI Proper Vertex Function of Gluon and the S-Matrix Element For $gg \rightarrow ggg$ Scattering Process at All Orders in Coupling Constant }

In this section we use eq. (\ref{5gk}) in the LSZ reduction formula to express the S-matrix element for the gluonic scattering process $gg \rightarrow ggg$ at all orders in coupling constant in QCD in terms of 5-point, 4-point, 3-point 1PI proper vertex functions and the (full) propagator by using the path integral formulation of QCD.

Consider the $2 \rightarrow 3$ gluonic scattering process $gg \rightarrow ggg$ in QCD given by
\bea
k_1+k_2 \rightarrow k'_1+k'_2+k'_3
\label{gs3}
\eea
where $k_1,k_2$ are the four-momenta of the two incoming gluons and $k'_1,k'_2,k'_3$ are the four-momenta of three outgoing gluons. The initial (final) state is given by
\bea
|i>=|k_1,k_2>,~~~~~~~~~~|f>=|k'_1,k'_2,k'_3>.
\label{ifi}
\eea
Using the LSZ reduction formula, the S-matrix element for the $gg \rightarrow ggg$ scattering process in eq. (\ref{gs3}) at all orders in coupling constant in QCD is given by
\bea
&&<f|i> = [G(-k'_1)]^{-1} [G(-k'_2)]^{-1}[G(-k'_3)]^{-1}[G(k_2)]^{-1}[G(k_1)]^{-1} G(-k'_1,-k'_2,-k'_3,k_1,k_2)\nonumber \\
\label{h3}
\eea
where $G(k)$ is the renormalized (full) propagator of gluon in momentum space and $G(k_1,k_2,k_3,k_4,k_5)$ is the renormalized 5-point connected green's function of gluon in momentum space. Note that as mentioned in eq. (\ref{cr}), for simplicity, we have included the finite factors such as the relevant sum over polarization vectors and color factors in the partonic cross section
\bea
{\hat \sigma} \propto |<f|i>|^2
\label{crf}
\eea
instead of the S-matrix element in eq. (\ref{h3}) so that the S-matrix element in eq. (\ref{h3}) is expressed in terms of the green's functions only.

By using eq. (\ref{5gk}) in (\ref{h3}) we find the expression of the S-matrix element for the scattering process $gg \rightarrow ggg$ at all orders in coupling constant in QCD in terms of 5-point, 4-point, 3-point 1PI proper vertex functions and the (full) propagator by using the path integral formulation of QCD.

The path integral procedure we have outlined above is suitable for the simultaneous study of renormalization of ultra violet (UV) divergences and factorization of infrared (IR) and collinear divergences at all orders in coupling constant in QCD.

The above technique can also be extended to the closed-time path integral formalism in non-equilibrium QCD \cite{c} to study partonic scattering cross sections at all orders in coupling constant in the non-equilibrium quark-gluon plasma \cite{a1,a2,a3,a4} at RHIC and LHC.

\section{Conclusions}
As far as the renormalization in perturbative QCD is concerned the n-point one particle irreducible (1PI) proper vertex function is the basic building block where the ultra-violet (UV) divergence occurs when the loop momentum integration limit goes to infinity. In this paper we have expressed the S-matrix element for the $gg \rightarrow ggg$ scattering process at all orders in coupling constant in terms of 5-point, 4-point, 3-point 1PI proper vertex functions and the (full) propagator by using the path integral formulation of QCD.

\end{document}